\documentclass{elsart3p}

\usepackage{graphicx}
\usepackage{graphics}
\usepackage{amssymb,amsmath}

\begin{document}

\begin{frontmatter}



\title{Theory of Josephson transport through spintronics nano-structure}


\author[label1,label2]{Shiro Kawabata}
\author[label3]{Yasuhiro Asano}
\author[label4]{Yukio Tanaka}
\author[label5]{Satoshi Kashiwaya}
\address[label1]{Nanotechnology Research Institute (NRI), National Institute of Advanced Industrial Science and Technology (AIST), Tsukuba, Ibaraki, 305-8568, Japan}
\address[label2]{CREST, Japan Science and Technology Corporation (JST), Kawaguchi, Saitama 332-0012, Japan}
\address[label3]{Department of Applied Physics, Hokkaido University, Sapporo, Japan}
\address[label4]{Department of Applied Physics, Nagoya University, Nagoya, Japan}
\address[label5]{Nanoelectronics Research Institute (NeRI), National Institute of Advanced Industrial Science and Technology (AIST), Tsukuba, Ibaraki, 305-8568, Japan}

\date{}

\begin{abstract}
We study the Josephson transport through ferromagnetic insulators  (FIs) by taking into account its band structure explicitly.
In the case of the fully polarized FIs (FPFIs), we found the formation of a $\pi$-junction and an atomic-scale 0-$\pi$ transition induced  by increasing the FI thickness.
More remarkably, in the Josephson junction through spin-filter materials such as Eu chalcogenides, the orbital hybridization between the conduction $d$  and the localized $f$ electron gives rise to the $\pi$-junction behavior.
Such FI-based $\pi$-junctions can be used to implement highly-coherent solid-state quantum bits.
\end{abstract}

\begin{keyword}
Josephson junction; Spintronics; Ferromagnetic insulator; Quantum bit
\end{keyword}
\end{frontmatter}

\section{Introduction}
\label{1}
The developing field of superconducting spintronics subsumes many fascinating physical phenomena with potential applications that may complement non-superconducting spintronics devices~\cite{rf:Zutic}.
 In addition there is an increasing interest in the novel properties of  junctions of superconductors and magnetic materials~\cite{rf:Golubov,rf:Buzdin1}.
One of the most interesting effects is the formation of a  Josephson $\pi$-junction in superconductor/ferromagnetic-metal/superconductor (S/FM/S) heterostructures~\cite{rf:Bulaevskii,rf:Buzdin2}.
In the ground-state phase difference between two coupled superconductors is $\pi$ instead of 0 as in the ordinary 0-junctions.
In terms of the Josephson relationship 
\begin{eqnarray}
I_J= I_C \sin \phi
,
\end{eqnarray}
where $\phi$ is the phase difference between the two superconductor layers, a transition from the 0 to $\pi$ states
implies a change in sign of the critical current $I_C$ from positive to negative. 
Physically, such a sign change of $I_C$ is a consequence of a phase change in the pairing wave-function induced in the FM layer due to the proximity effect.
The existence of the $\pi$-junction in S/FM/S systems has been confirmed in experiment by Ryanzanov et al.\cite{rf:Ryanzanov} and Kontos et al.\cite{rf:Kontos}.

Recently, quiet qubits consisting of a superconducting loop with a S/FM/S $\pi$-junction have been proposed~\cite{rf:Ioffe,rf:Blatter}.
In quiet qubits, a quantum two-level system  is spontaneously generated and therefore it is expected to be robust to the decoherence by the fluctuation of the external magnetic field.
From the viewpoint of the quantum dissipation, however,  S/FM/S junctions are inherently identical with S/N/S junctions (N is a normal nonmagnetic metal).
Thus a gapless quasiparticle excitation in the FM layer is inevitable.
This feature gives a strong dissipative or decoherence effect~\cite{rf:Schon}.
Therefore the realization of the $\pi$-junction $without$ a $metallic$ interlayer is highly desired for qubit applications~\cite{rf:Kawabata1,rf:Kawabata2,rf:Kawabata3,rf:Kawabata4}.

In this paper, we investigate the Josephson effect through a ferromagnetic $insulators$ (FIs) numerically.
Although the $\pi$-junction formation in such junctions has been theoretically predicted~\cite{rf:Tanaka} and subsequently analyzed by use of the quasiclassical Green's function techniques~\cite{rf:Fogelstrom,rf:Zhao}, a phenomenological $\delta$-function potential have been used in order to model the FI barrier.
Then a natural question to ask is can we realize  the $\pi$-junction in $actual$ FIs?
Moreover the possibility of the $\pi$-junction formation in the $finite$ FI-barrier $thickness$ case  is also an interesting and unresolved problem.
In order to resolve above issues, we will formulate a numerical method for the Josephson current through FIs by taking into account the band structure and the finite thickness of FIs explicitly~\cite{rf:Kawabata5}.
Then we will show the possibility of the  $\pi$-junction formation for  two important representative  FIs in the  spintronics field~\cite{rf:Nagahama,rf:Felser}, i.e., the fully polarized FI (FPFI)  (e.g., La${}_2$BaCuO${}_5$) and the spin-filter materials  (e.g., Eu chalcogenides).

\section{Model}
\label{2}

Let us consider a two-dimensional tight-binding lattice of the S/FI/S junctions as shown in Fig.~1(a).
The vector 
\begin{eqnarray}
{\bf r}=j{\bf x}+m{\bf y}
\end{eqnarray}
 points to a lattice site, where ${\bf x}$ and ${\bf y}$ are unit vectors in the $x$ and $y$ directions,
respectively.
In the $y$ direction, we apply the periodic boundary condition for the number of lattice sites being $W$.
Electronic states in a superconductor are described by the
mean-field Hamiltonian, 
\begin{eqnarray}
H_\mathrm{BCS}
&=&
\frac{1}{2} \sum_{{\bf r},{\bf r}^{\prime }  \in \mathrm{S}    }%
\left( \tilde{c}_{{\bf r}}^{\dagger }\;\hat{h}_{{\bf r},%
{\bf r}^{\prime }}\;\tilde{c}_{{\bf r}^{\prime }}^{{}}-%
\overline{\tilde{c}_{{\bf r}}}\;\hat{h}_{{\bf r},{\bf r}%
^{\prime }}^{\ast }\;\overline{\tilde{c}_{{\bf r}^{\prime }}^{\dagger
}}
\right)
\nonumber\\
&+&
\frac{1}{2} 
\sum_{{\bf r}\in \mathrm{S}} 
\left( \tilde{c}_{%
{\bf r}}^{\dagger }\;\hat{\Delta}\;\overline{\tilde{c}_{{\bf r}
}^{\dagger }}-\overline{\tilde{c}_{{\bf r}}}\;\hat{\Delta}^{\ast }\;%
\tilde{c}_{{\bf r}} 
\right)  
.
\end{eqnarray}
Here
\begin{eqnarray}
\hat{h}_{{\bf r},{\bf r}^{\prime }}=
\left[ -t_s \delta _{|%
{\bf r}-{\bf r}^{\prime }|,1}
+
(-\mu_s+4t_s)\delta _{{\bf r},{\bf r}^{\prime }}
\right] \hat{\sigma}_{0}
,
\end{eqnarray}
with 
\begin{eqnarray}
\overline{\tilde{c}}_{{\bf r}}=\left( c_{{\bf r}%
,\uparrow },c_{{\bf r},\downarrow }\right) ,
\end{eqnarray}
 where
  $
  c_{{\bf r} ,\sigma }^{\dagger }$ ($c_{{\bf r},\sigma }^{{}}
$)
 is the creation
(annihilation) operator of an electron at ${\bf r}$ with spin
 $\sigma
=$ ( $\uparrow $ or $\downarrow $ ), $\overline{\tilde{c}}$ 
means the
transpose of $\tilde{c}$,  and $\hat{\sigma}_{0}$ is $2\times 2$ unit matrix. 
The chemical potential  $\mu_s$ is set to be $2 t_s$ for superconductors.
In superconductors, the hopping integral $t_s$ is considered among nearest neighbor sites and we choose 
\begin{eqnarray}
\hat{\Delta}=i\Delta \hat{\sigma}_{2},
\end{eqnarray}
 where $\Delta $ is the amplitude 
of the pair potential in the $s$-wave symmetry channel, and $\hat{\sigma}_{2}$ is a Pauli matrix.

\begin{figure}[t]
\begin{center}
\includegraphics[width=7.6cm]{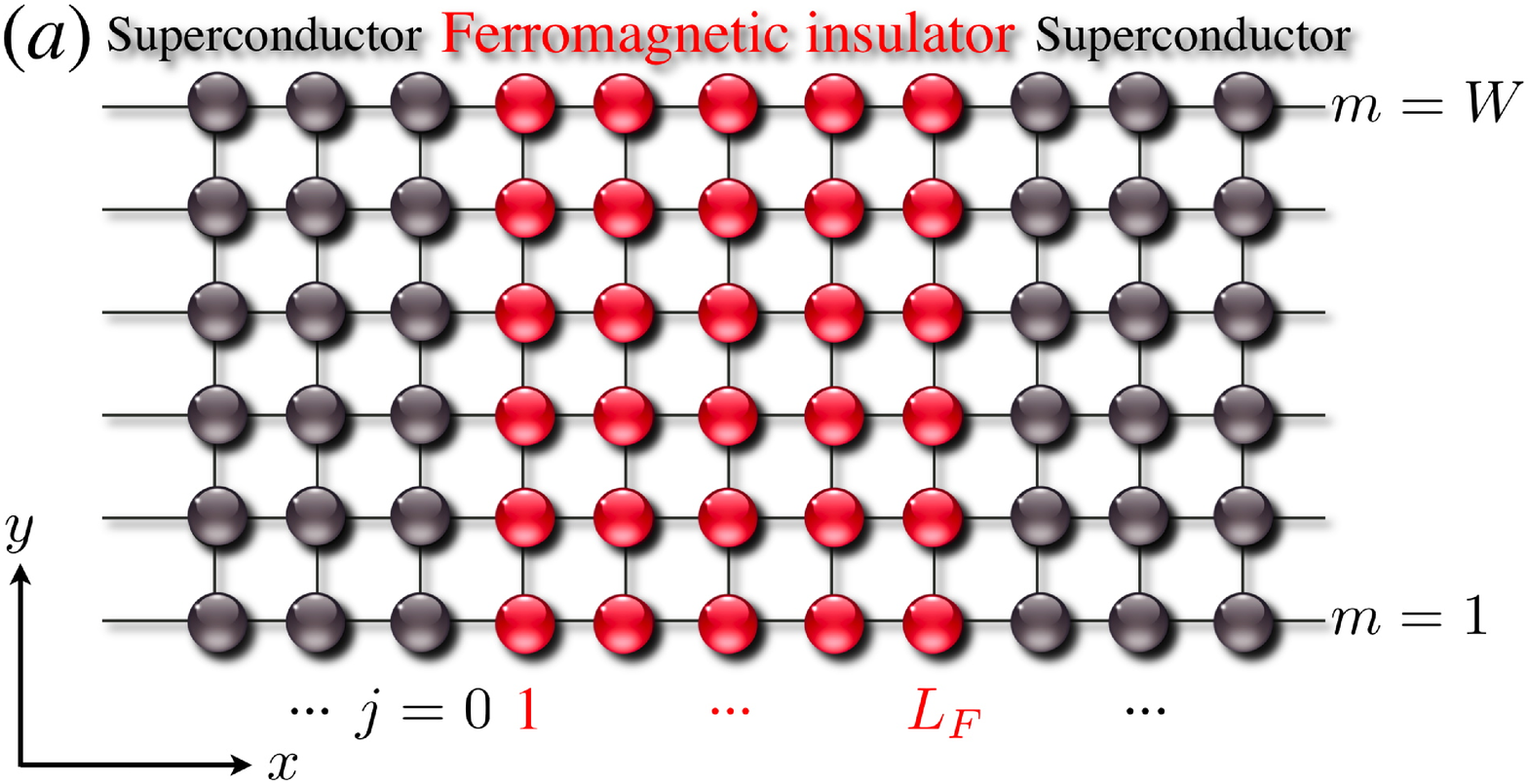}
\includegraphics[width=7.4cm]{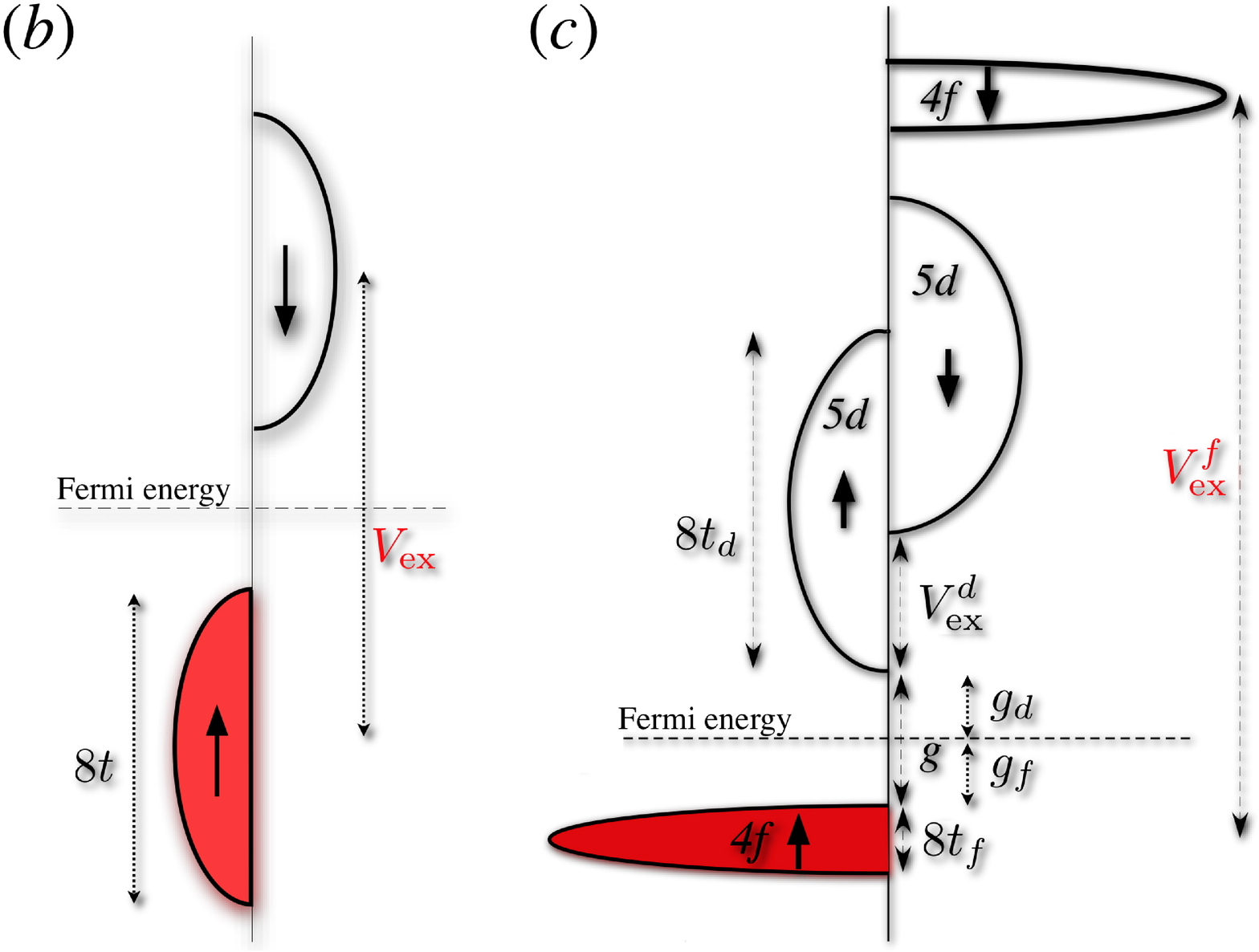}
\end{center}
\caption{(Color online) (a) A schematic figure of a Josephson junction through the ferromagnetic insulators on the
tight-binding lattice. The density of states for each spin direction for (b) the fully polarized ferromagnetic insulator, e.g., LBCO, and (c) the spin-filter materials, e.g., Eu-chalcogenides.
 }
\label{fig1}
\end{figure}

We consider two representative FIs as a barrier of the Josephson junction, i.e., FPFIs [Fig. 1(b)]  and spin-filter materials [Fig. 1(c)].
The typical density of states (DOS) of FPFIs for each spin direction is shown schematically in Fig. 1(b).
One of the important FPFIs is La${}_2$BaCuO${}_5$ (LBCO)~\cite{rf:Mizuno,rf:Ku}.
The exchange splitting $V_\mathrm{ex}$ is estimated  to be 0.34 eV by a first-principle band calculation~\cite{rf:LBCO}.
Since $V_\mathrm{ex}$ is large and the bands are originally half-filled, the system becomes FI. 

The Hamiltonian of FPFI layer is described by a single-band tight-binding model as 
\begin{eqnarray}
H_\mathrm{FPFI}
& =& -t \sum_{{\bf r},{\bf r}^{\prime },\sigma} 
c_{{\bf r},\sigma}^\dagger 
c_{{\bf r}',\sigma}
-\sum_{{\bf r}} ( 4 t -\mu)  
c_{{\bf r},\uparrow}^\dagger 
c_{{\bf r},\uparrow}
\nonumber\\
&+&
 \sum_{{\bf r}} 
( 4 t -\mu + V_\mathrm{ex}) 
 c_{{\bf r},\downarrow}^\dagger 
 c_{{\bf r},\downarrow}
,
\end{eqnarray}
where $V_\mathrm{ex}$ is the exchange splitting [see Fig. 1(b)].
If $V_\mathrm{ex} >  8 t$ ($V_\mathrm{ex} <  8 t$), this Hamiltonian describes FPFI (FM).
The chemical potential $\mu$ is given by
\begin{eqnarray}
 \mu= \frac{V_\mathrm{ex}}{2}  + 4t
 . 
\end{eqnarray}

Recently a spin-filter effect has been intensively studied for spintronics applications~\cite{rf:Nagahama}.
Typical spin-filter material is  a Eu chalcogenide, e.g., EuO and EuS.
 The schematic DOS of  the Eu chalcogenides is shown in Fig. 1(c).
The Eu chalcogenides stand out among the FIs as ideal Heisenberg ferromagnets, with a high magnetic moment and a large exchange splitting of the conduction band for Eu $5d$-electrons.
Ferromagnetic order of the $4f$ spins causes exchange splitting of the conduction 5$d$ band, lowering (raising) the spin-up (-down) band symmetrically by $V_\mathrm{ex}^d/2$.

For the spin-filter materials such as Eu chalcogenides, we use a following $d$-$f$ Hamiltonian including the $d$-$f$ hybridization~\cite{rf:d-f1,rf:d-f2},
\begin{eqnarray}
H_\mathrm{SF} = H_d + H_f +H_{df}.
\end{eqnarray}
Here
\begin{eqnarray}H_d
& = &
-t_d \sum_{{\bf r},{\bf r}^{\prime },\sigma} 
d_{{\bf r},\sigma}^\dagger 
d_{{\bf r}',\sigma}
+\sum_{{\bf r}} ( 4 t_d -\mu_d)  
d_{{\bf r},\uparrow}^\dagger 
d_{{\bf r},\uparrow}
\nonumber\\
&+& \sum_{{\bf r}} 
( 4 t_d -\mu_d + V_\mathrm{ex}^d) 
 d_{{\bf r},\downarrow}^\dagger 
 d_{{\bf r},\downarrow}
 ,
 \end{eqnarray}
\begin{eqnarray}
H_f 
& = &
-t_f \sum_{{\bf r},{\bf r}^{\prime },\sigma} 
f_{{\bf r},\sigma}^\dagger 
f_{{\bf r}',\sigma}
+\sum_{{\bf r}}
 ( 4 t_f -\mu_f) 
  f_{{\bf r},\uparrow}^\dagger 
  f_{{\bf r},\uparrow}
\nonumber\\
&+&
 \sum_{{\bf r}} 
( 4 t_f -\mu_f + V_\mathrm{ex}^f)  
f_{{\bf r},\downarrow}^\dagger
 f_{{\bf r},\downarrow}
 ,
 \end{eqnarray}
\begin{eqnarray}
  H_{df}
=
 V_{df} \sum_{{\bf r},\sigma} 
 \left(
 d_{{\bf r},\sigma}^\dagger  f_{{\bf r},\sigma}
 +
  f_{{\bf r},\sigma}^\dagger  d_{{\bf r},\sigma}
 \right)
 ,
 \end{eqnarray}
where  $d_{{\bf r} ,\sigma }^{\dagger }$ $(f_{{\bf r} ,\sigma }^{\dagger }$) is the creation operator, $t_d$ $(t_f)$ is the hopping integral
and  $V_\mathrm{ex}^d$ $(V_\mathrm{ex}^f)$ is the exchange splitting of $d$ $(f)$ electrons.
The chemical potential of $d$ and $f$ electrons is respectively given by 
\begin{eqnarray}
\mu_d &=& -g_d
\\
\mu_f &=& 8 t_f + g_f
 ,
 \end{eqnarray}
 where $g_d$ $(g_f)$ is the energy gap of the $d$ $(f)$ band [see Fig. 1(c)].
The third term $H_{df}$ of the Hamiltonian describes the mixing between $d$ and $f$ electrons.

The Hamiltonian is diagonalized by the Bogoliubov transformation and the
Bogoliubov-de Gennes equation is numerically solved by the recursive
Green function method~\cite{rf:Asano1}.
We calculate the Matsubara
Green function $\check{G}_{\omega _{n}}({\bf r},{\bf r}^{\prime })$ in a FI layer, 
\begin{equation}
\check{G}_{\omega _{n}}({\bf r},{\bf r}^{\prime })=\left[
\begin{array}{cc}
\hat{g}_{\omega _{n}}({\bf r},{\bf r}^{\prime }) & \hat{f}%
_{\omega _{n}}({\bf r},{\bf r}^{\prime }) \\
-\hat{f}_{\omega _{n}}^{\ast }({\bf r},{\bf r}^{\prime }) & -%
\hat{g}_{\omega _{n}}^{\ast }({\bf r},{\bf r}^{\prime })%
\end{array}
\right] , \label{deff}
\end{equation}
where 
\begin{equation}
\omega _{n}= (2 n + 1) \pi T
\end{equation}
 is the Matsubara frequency.
The Josephson current is given by
\begin{equation}
I_J (\phi)=-ietT\sum_{\omega _{n}}\sum_{m=1}^{W}\mathrm{Tr}\left[ \check{G}_{\omega
_{n}}({\bf r}^{\prime },{\bf r})-\check{G}_{\omega _{n}}(%
{\bf r},{\bf r}^{\prime })\right]
,
\end{equation}
with $T$ being a temperature and ${\bf r}^{\prime }={\bf r}+{\bf x}$. 
Throughout this paper we fix the following
parameters: $W=25$, and $\Delta _{0}=0.01t$, $T=0.01T_{c}$, where $T_c$ is the superconductor transition temperature.

\section{Josephson current through FPFI}
\label{3}

\begin{figure}[t]
\begin{center}
\includegraphics[width=7.5cm]{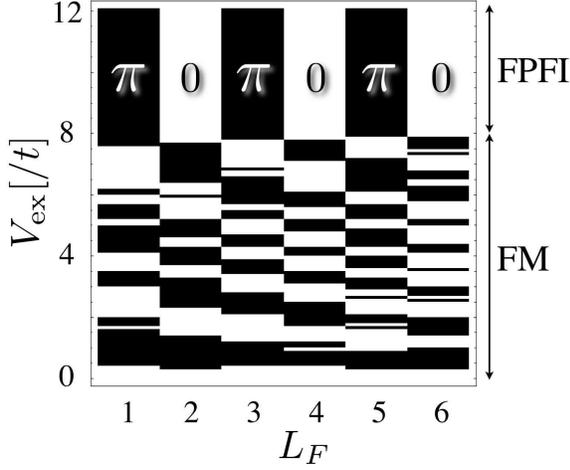}
\end{center}
\caption{The 0-$\pi$ phase diagram depending on the exchange splitting $V_\mathrm{ex}$ and the thickness $L_F$ for FM ($0 \le V_\mathrm{ex}/t  \le 8$) and the fully polarized FI ($V_\mathrm{ex}/t > 8$). 
The black and white regime correspond to the $\pi$- and 0-junction, respectively.}
\label{fig2}
\end{figure}

 We first investigate the Josephson transport through FPFIs [Fig. 1(b)].
In the calculation, we assume $t=t_s$ for simplicity.
The phase diagram depending on the strength of $V_\mathrm{ex}$ ($0 \le V_\mathrm{ex}/t  \le 8$ for FM and $V_\mathrm{ex}/t > 8$ for FPFI) and $L_F$ is shown in Fig.~\ref{fig2}.
The black (white) regime corresponds to the $\pi$- (0-)junction, i.e., $I_J=  -(+)I_C \sin \phi$.
In the case of FPFI, the $\pi$-junction can be formed.
Moreover remarkably, {\it the atomic-scale 0-$\pi$ transition is induced  by increasing the thickness of the FI barrier $L_F$.}
The physical origin of the 0-$\pi$ transition will be discussed in elsewhere~\cite{rf:Kawabata6}.
%
%
%
%
%
%

\section{Josephson current through spin-filter materials}
\label{4}

Next we consider the Josephson transport through the spin-filter materials such as the Eu-chalcogenides.
In the calculation, we use the following parameters in consideration of EuO: $t_d=1.25$ eV, $g_d =7.5$ eV, $g_f =0.075$ eV, and $V_\mathrm{ex}^d=0.528$ eV.
For simplicity, we assume $t_d=t_s$.

We first discuss the Josephson current through the $5d$-band only in order to check whether the spin-filter effect gives rise to the $\pi$-junction behavior or not.
In this case, we numerically found that no $\pi$-junction is formed irrespective of the thickness $L_F$ and $V_\mathrm{ex}^d$.
Therefore the spin-filter effect gives rise to only the $0$-junction behavior.

Next we consider the Josephson transport through the Eu-chalcogenides including both the $5d-$ and $4f$-band.
In the calculation we set $L_F=5$.
We  systematically change the values of the hopping integral for the $f$ band $t_f$ ($=0.0\sim 0.5 t_d$) and the $d$-$f$ 
hybridization $V_{df}$ $(= 0 \sim 6.5 t_d$ ).
Fig. 3 shows the 0-$\pi$ phase diagram numerically obtained.
Remarkably, the $\pi$-junction is realized at the certain values of $t_f$ and $V_{df}$.
We found that the $\pi$-junction can be formed if (1) $d$ and $f$ band for down spin are overlapped each other (see inset in Fig. 3) and (2) {\it  the $d$-$f$ hybridization $V_{df}$ is strong enough}.
More detailed discussion for above results will be given in elsewhere~\cite{rf:Kawabata6}.

\section{Summary}
\label{5}

To summarize, we have studied the Josephson effect in S/FI/S junction by use of the recursive Green's function method.
We found that the $\pi$-junction and the atomic-scale 0-$\pi$ transition is realized in the case of FPFIs.
On the other hand, in the case of the Josephson junction with the spin-filter material, the $\pi$-junction can be formed if the $d$ and $f$ bands are overlapped and  the $d$-$f$ hybridization is strong. 
Such FI based $\pi$-junctions can be used as an  element in the architecture of  {\it ideal quiet qubits} which possess both the quietness and the weak quasiparticle-dissipation nature.
Therefore, ultimately, we could realize a FI-based highly-coherent quantum computer.

We  would like to thank J. Arts, A. Brinkman, M. Fogelstr\"om, A. A. Golubov, P. J. Kelly, T. L\"ofwander, T. Nagahama, F. Nori, J. Pfeiffer, and M. Weides for useful discussions.
This work was  supported by CREST-JST, and a Grant-in-Aid for Scientific Research from the Ministry of Education, Science, Sports and Culture of Japan (Grant No. 19710085).

\begin{figure}[tb]
\begin{center}
\includegraphics[width=7.0cm]{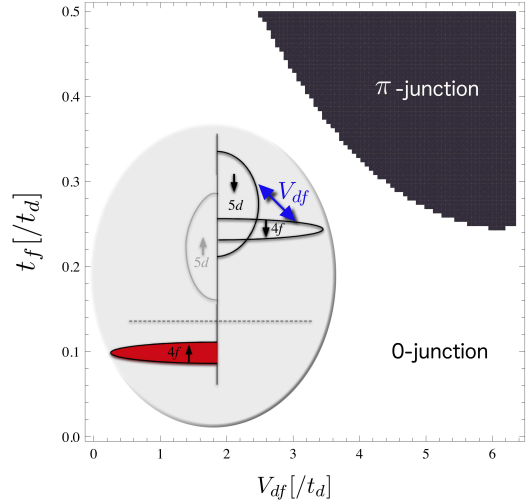}
\end{center}
\caption{(Color online) The 0-$\pi$ phase diagram depending on the $d$-$f$ hybridization $V_{df}$ and the hopping integral $t_f$ of the $f$ electrons for the Josephson junction through the spin-filter materials.
The black and white regime correspond to the $\pi$- and 0-junction. 
Inset shows a schematic DOS configuration for the junction.}
\label{fig3}
\end{figure}
%
%
%
%
%


\end{document}